\documentclass{WileyMSP-template}
\usepackage{xcolor}
\begin{document}
\pagestyle{fancy}
\rhead{\includegraphics[width=2.5cm]{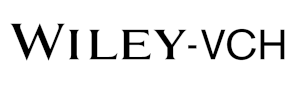}}

\newcommand{\pd}[1]{#1}
\title{Observation of relativistic domain wall motion in amorphous ferrimagnets}

\maketitle


\author{Pietro Diona*}
\author{Luca Maranzana}
\author{Sergey Artyukhin*}
\author{Giacomo Sala*}


\begin{affiliations}
Pietro Diona\\
Nanoscience, Scuola Normale Superiore, Piazza dei Cavalieri 7, Pisa, Italy\\
Quantum Materials Theory, Italian Institute of Technology, Via Morego 30, Genoa, Italy\\
Department of Materials, ETH Zürich, Hönggerbergring 64, 8093, Zürich, Switzerland\\
pietro.diona@sns.it\\

Luca Maranzana\\
Quantum Materials Theory, Italian Institute of Technology, Via Morego 30, Genoa, Italy\\
Department of Physics, University of Genoa, Via Dodecaneso 33, Genoa, Italy\\
luca.maranzana@iit.it\\

Dr. Sergey Artyukhin\\
Quantum Materials Theory, Italian Institute of Technology, Via Morego 30, Genoa, Italy\\
sergey.artyukhin@iit.it\\

Dr. Giacomo Sala\\
Department of Quantum Matter Physics, University of Geneva, Quai Ernest Ansermet 24,\\
Geneva, Switzerland\\
Department of Materials, ETH Zürich, Hönggerbergring 64, 8093 Zürich, Switzerland\\
giacomo.sala@unige.ch
\end{affiliations}



\begin{abstract}
Domain walls in ferrimagnets and antiferromagnets behave as relativistic sine-Gordon solitons with \pd{the spin-wave group velocity setting the ultimate velocity of domain walls and speed of magnetic devices}. While this relativistic regime has been achieved in crystalline ferrimagnets, they cannot be routinely integrated in devices. To enable technological \pd{breakthroughs}, relativistic dynamics must be demonstrated in easy-to-integrate ferrimagnets \pd{such as rare-earth -- transition-metal alloys. However, this scenario} remains elusive due to the inherent magnetic disorder of these materials, \pd{their} complex spin-wave spectra, and challenges in modeling their ultrafast dynamics. Here, we demonstrate relativistic domain wall motion in amorphous ferrimagnetic GdFeCo \pd{devices operated in the proximity of the angular momentum compensation point}. The current-induced domain wall velocity saturates within 10\% of the spin-wave speed of 2 km/s\pd{, a behavior consistent with relativistic model of domain wall motion}. Our observation of relativistic dynamics in technologically relevant ferrimagnets opens the way to magnetic devices operating at the ultimate speed limit.
\end{abstract}


\section{Introduction}
Magnetic domain walls have been proposed as fundamental units of racetrack memories for ultra-dense storage and logic applications \cite{parkin2008magnetic}. Domain walls were initially manipulated using the current-induced spin transfer torque, a transfer of angular momentum between localized and conduction electron spins mediated by the exchange interaction in a magnetic thin film \cite{STT,brataas2012current}. A more efficient control of domain wall dynamics was later achieved with spin-orbit torques (SOT). In this case, angular momentum is generated by the spin Hall and/or Rashba-Edelstein effect in an adjacent nonmagnetic material or at the magnet-nonmagnet interface \cite{manchon2019current, ryu2020current}. This effect provides a more versatile and faster way to control magnetization dynamics \cite{manchon2019current,SOT1, diona2022simulation, ramaswamy2018recent}. 

Diverse magnetic materials have been explored to optimize the trade-off between energy consumption and performance of magnetic devices based on SOT-driven domain walls \cite{shao2021roadmap}. While early studies focused on ferromagnets, attention has recently shifted to materials that offer higher energy efficiency and domain wall velocities, such as antiferromagnets \cite{baldrati2019mechanism, gomonay2016high, baltz2018antiferromagnetic, shiino2016antiferromagnetic} and ferrimagnets \cite{caretta2024domain, kim2022ferrimagnetic, vsteady1}. In ferromagnets, when an applied  magnetic field exceeds the Walker breakdown threshold, excessive torque on the domain wall induces precessional motion \cite{beach2005dynamics, mougin2007domain}, which leads to a velocity drop. This effect vanishes in compensated magnetic materials such as antiferromagnets \cite{baltz2018antiferromagnetic}, which thus represent a promising alternative to ferromagnets. However, antiferromagnetic dynamics are notoriously difficult to control and probe. Ferrimagnets offer instead a good compromise between ferro- and antiferromagnetic systems. Many ferrimagnetic materials can be easily studied with conventional magnetic techniques, like ferromagnets, but exhibit low sensitivity to external magnetic fields, as antiferromagnets \cite{kim2022ferrimagnetic}. Importantly, they also support ultrafast domain wall dynamics due to the suppression of the Walker breakdown, which makes them interesting candidates for realistic magnetic applications \cite{zhang2023ferrimagnets}.
Among these materials, rare-earth -- transition-metal (RE-TM) ferrimagnets stand out thanks to their exceptional tunability and easy integration into devices \cite{finley2020spintronics, kim2022ferrimagnetic, sala2022ferrimagnetic}. \pd{By adjusting the composition, it is possible to easily and continuously} tune their anisotropy, net magnetization, and net angular momentum, which is crucial for achieving domain wall speeds in the order of several km/s \cite{caretta2018fast, cai2020ultrafast, kim2017fast, sala2022asynchronous}. \pd{Moreover, their amorphous and metallic nature and compatibility with the multilayered thin-film structures used in magnetic tunnel junctions and racetrack memories \cite{gonzalez2021} enable scalable deposition on virtually any substrate and integration into large-scale device production.}

\pd{The high speeds typical of these materials} can drive domain walls into nonclassical regimes where conventional models of domain wall dynamics cease to be valid. In this scenario, domain walls are best described as sine-Gordon solitons with relativistic kinematics \cite{sinegordon1}. As the domain wall approaches the maximum spin-wave group velocity, its width contracts and spin waves are emitted \cite{tatara2020magnon, gomonay2016high, yan2011fast}. The maximum spin-wave group velocity sets, therefore, an upper bound on the domain wall speed, similar to the speed of light in the theory of special relativity. This limit determines the highest operational frequency of magnetic devices and is thus an important parameter to identify. Moreover, as the emitted spin waves are in the THz range, these relativistic effects can serve as THz sources of electric signals. Relativistic kinematics has been observed in crystalline yttrium-iron garnet ferrimagnets \cite{caretta1}. \pd{However, these epitaxial materials} are difficult to synthesize and integrate in devices \pd{because of constraints such as the high-temperature growth, challenging precise control of the stoichiometry, and limited substrate compatibility. Moreover, the insulating character of garnets hinders the use of conventional transport probes such as the anomalous Hall effect and tunneling magnetoresistance. At a fundamental level,} relativistic kinematics is expected on general grounds in other magnetic materials \cite{alliati2022relativistic} and, in particular, has been predicted for RE-TM ferrimagnets with ultrafast domain wall speeds and strong technological potential \cite{spinwave}. Yet, evidence for relativistic effects in materials other than ferrimagnetic garnets is still lacking.

Here, we show that SOT can drive domain walls up to relativistic speeds in ferrimagnetic GdFeCo alloys. We observe a saturation of the domain wall speed to \(\approx 1.7\) km/s independently of the applied magnetic field. This saturation cannot be explained by classical models of domain wall dynamics but is captured by a relativistic model with a maximum spin-wave group velocity of \(\approx 2\) km/s. \pd{These observations establish the possibility of achieving the relativistic regime in a vast class of magnetic alloys and set the stage for efficient spintronic and THz applications.}

\section{Modeling relativistic dynamics in RE-TM ferrimagnets}
We first describe how the maximum spin-wave group velocity limits the domain wall speed. We consider a metallic heterostructure composed of a ferrimagnetic layer and a heavy-metal film, which sources the SOT. Domain walls in the ferrimagnet are driven by the combined action of the dampinglike torque field $\tilde{H}_\mathrm{DL}$ and the external magnetic field  $H_\mathrm{x} = \frac{B_\mathrm{x}}{\mu_0}$ applied in the sample plane parallel to the current direction x \cite{manchon2019current}. The magnetic field is necessary to transform Bloch walls, which are typical of materials with low Dzyaloshinskii-Moriya interaction (DMI) and cannot be driven by SOT \cite{khvalkovskiy2013matching}, into Néel domain walls. The steady-state motion of the domain wall is then described by a relativistic equation that predicts a velocity $v_\mathrm{DW}$ \cite{caretta1}:
\begin{equation}
\label{rel2}
    v_\mathrm{DW} = \frac{v_\mathrm{g, max}}{\sqrt{1 + \left(\frac{v_\mathrm{g, max}}{v_\mathrm{cl}}\right)^{2}}}.
\end{equation}
Here, $v_\mathrm{g, max} = \frac{2A}{ds_\mathrm{T}}$ is the maximum spin-wave group velocity, which plays the same role as the speed of light in the theory of special relativity \cite{fogel1977dynamics, huang2015exact}. It depends on the antiferromagnetic exchange interaction $A$, the average interatomic distance $d$, and the total spin density $s_\mathrm{T} = s_\mathrm{A} + s_\mathrm{B}$ of the two sublattices composing the ferrimagnet. $v_\mathrm{cl}$ is the classical domain wall velocity far away from $v_\mathrm{g, max}$ and reads \cite{haltz2021domain}:

\begin{eqnarray}
\label{vsteadycomplete}
    &v_\mathrm{cl} = \frac{\Delta\gamma_\mathrm{eff}\tilde{H}_\mathrm{pin}}{\alpha_\mathrm{eff}\beta}\left[1 - \sqrt{1 - \beta\left[1 - \left(\frac{2\eta\tilde{H}_\mathrm{DL}}{\tilde{H}_\mathrm{pin}}\right)^2\right]}\right],\nonumber\\
    & \beta = 1 + \left[ \frac{\frac{\pi}{2}\tilde{H}_\mathrm{DL}}{\alpha_\mathrm{eff}\left(H_\mathrm{D} + \frac{\pi}{2}H_\mathrm{x}\right)}\right]^{2},
\end{eqnarray}

\noindent where $\Delta \sim \sqrt{\frac{A}{K_{\mathrm{eff}}}}$ is the domain wall width, $K_{\mathrm{eff}}$ is the effective anisotropy, $\alpha_\mathrm{eff} = \frac{\alpha s_\mathrm{T}}{\delta_\mathrm{s}}$ is the effective Gilbert damping parameter of the system, \(\delta_\mathrm{s} = s_\mathrm{A} - s_\mathrm{B}\) is the net angular momentum, $\gamma_\mathrm{eff}$ is the effective gyromagnetic ratio, and $H_\mathrm{D}$ is the DMI field. The damping like spin-orbit field $\tilde{H}_\mathrm{DL} \sim \frac{\theta_{\mathrm{DL}} j}{M_{\mathrm{S}}}$ is proportional to the current density $j$ and dampinglike SOT efficiency $\theta_{\mathrm{DL}}$ and is inversely proportional to the saturation magnetization $M_{\mathrm{S}}$. We also consider with $\eta\leq 1$ the domain wall broadening caused by the in-plane magnetic field, which reduces the effective action of the SOT, and pinning effects described by the effective field $\tilde{H}_\mathrm{pin}$ \cite{jiang2010enhanced} (Supporting Information). 

\begin{figure}[b]
\includegraphics[width=\linewidth]{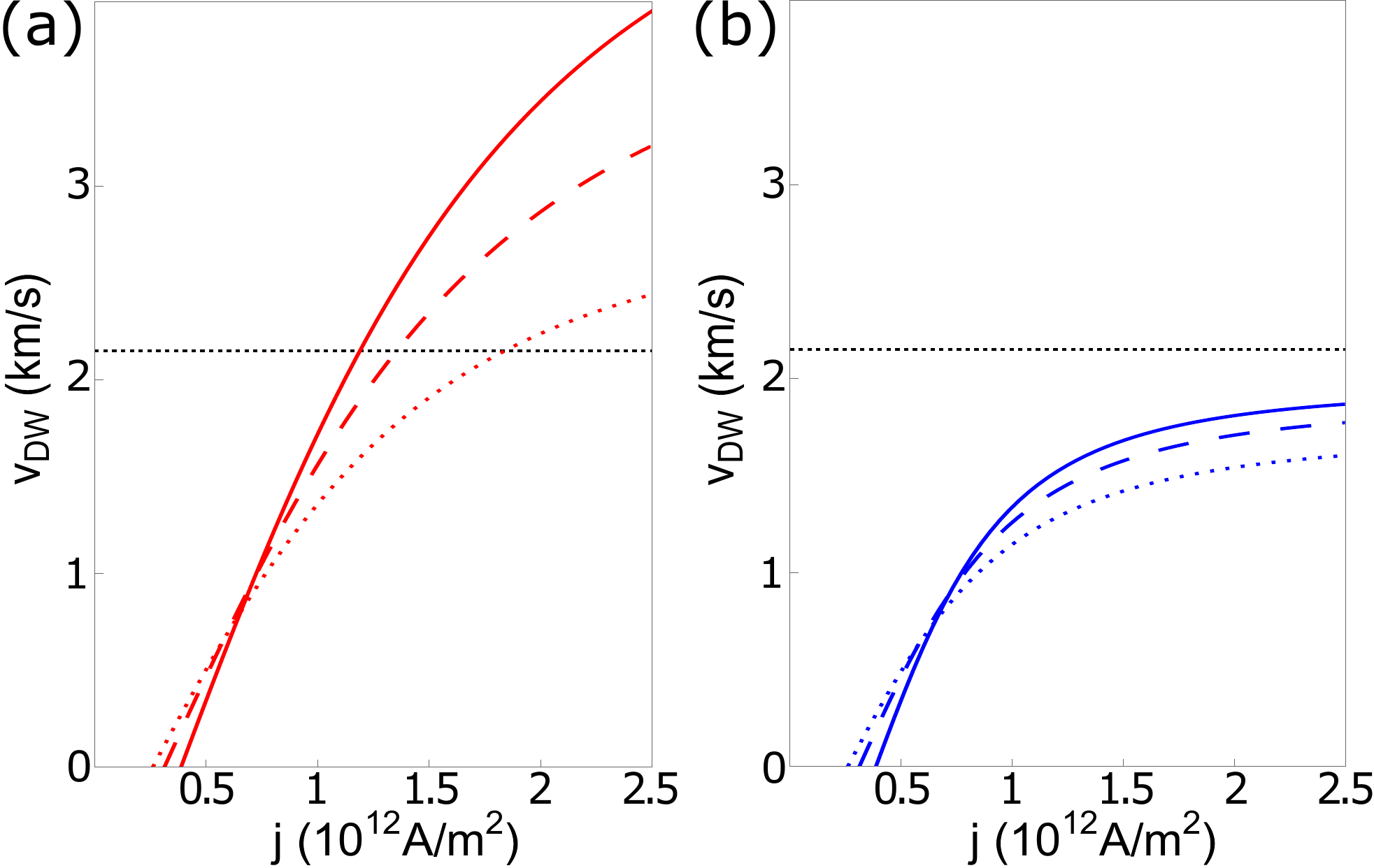}
     \caption{(a) Classical (Equation~\ref{vsteadycomplete}) and (b) relativistic (Equation~\ref{rel2}) domain wall velocity as a function of the current density for different in-plane magnetic fields: $B_\mathrm{x}$ = 200 mT (continuous line), 150 mT (dashed line), 100 mT (dotted line). The relativistic model predicts a strong saturation close to the spin-wave group velocity (black dashed line). The parameters used for the calculation are taken from Table \ref{paramextracted}.}
     \label{comparison} 
\end{figure}

According to Equation~\ref{vsteadycomplete}, the classical steady state speed $v_\mathrm{cl}$ results from the competition between opposite wall deformations caused by the external field $H_\mathrm{x}$ and the current-induced SOT field $\tilde{H}_\mathrm{DL}$. While $H_\mathrm{x}$ and the DMI $H_\mathrm{D}$ keep the domain wall magnetization parallel to the electric current, $\tilde{H}_\mathrm{DL}$ tends to rotate it towards the transverse in-plane direction. $H_\mathrm{x}$ and $H_\mathrm{D}$ counteract, therefore, the tendency of the dampinglike field to transform the Néel domain wall into a Bloch wall, which is detrimental to the efficient domain wall drive by SOT \cite{khvalkovskiy2013matching}. As a consequence, this interplay leads to a saturation of $v_\mathrm{cl}$ at high current densities when $H_\mathrm{x}$ becomes incapable of counterbalancing the SOT (\textbf{Figure \ref{comparison}a}). Yet, increasing the strength of $H_\mathrm{x}$ allows \pd{for pushing the saturation velocity to arbitrarily high values}, i.e., the classical model does not set any limits to the maximum domain wall speed, provided that the restoring torque in the domain wall can be adjusted accordingly. This nonphysical result is redeemed by the relativistic model in Equation~\ref{rel2}. As $H_\mathrm{x}$ increases, the saturation velocity increases asymptotically towards the maximum spin-wave velocity and never exceeds it (\textbf{Figure \ref{comparison}b}). Thus, the relativistic domain wall dynamics are characterized by two sources of saturation: the competition between the SOT and in-plane field in the low-field regime (Equation ~\ref{vsteadycomplete}) and the spin-wave group velocity in the high-current, high-field regime (Equation~\ref{rel2}).

These relativistic dynamics have been observed in crystalline iron-garnet ferrimagnets with velocities $\approx 4.3$ km/s \cite{caretta1}. However, the model \pd{in Equation~\ref{rel2}} is generally applicable to any ferrimagnetic materials, and a relativistic domain wall motion is expected in other systems. In RE-TM ferrimagnets, \(v_{\mathrm{g,max}}\) was first deemed experimentally unattainable \cite{caretta1} due to the large antiferromagnetic exchange interaction $A$ in these materials, if compared to ferrimagnetic garnets. However, $A$ depends sensitively on the stoichiometry of the alloy. For values \(A \approx 1-2~\mathrm{pJ/m}\) typical of GdFeCo alloys \cite{raasch1994exchange}, we expect \(v_{\mathrm{g,max}} \approx 1.5-2~\mathrm{km/s}\), which can be measured with conventional probes of magnetic dynamics. These considerations, therefore, suggest that RE-TM ferrimagnets can also exhibit relativistic motion in response to SOT and magnetic fields.

\section{Observation of relativistic domain wall motion}
\begin{figure*}[t]
     \centering
    \includegraphics[width = 1\textwidth]{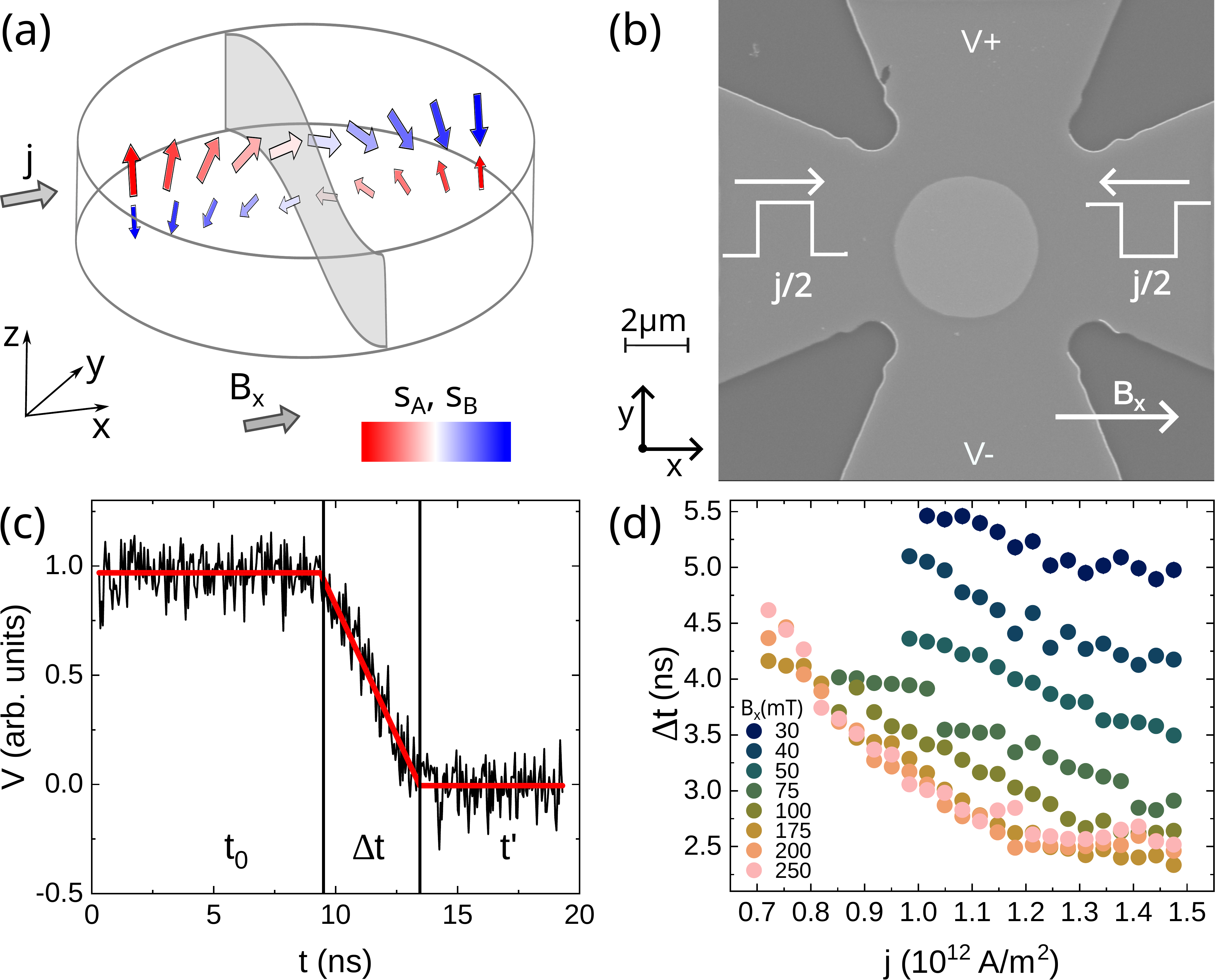}
    \caption{(a) A sketch of a domain wall inside the dot driven by the electric current density $j$ and in-plane magnetic field $B_\mathrm{x}$. The upper and lower arrows indicate the spins $S_\mathrm{A, B}$ of the two sublattices A and B. (b) Electron microscopy image of the device and sketch of the experimental setup used to measure the domain wall motion across a GdFeCo dot. $j$ is the current density and $V = V_+ - V_-$ is the Hall voltage. (c) Exemplary switching trace of a 4 \textmu m-wide Gd$_{30}$Fe$_{63}$Co$_7$ dot, measured while injecting current pulses of amplitude \(1.15\cdot 10^{12}~\mathrm{A/m^2}\) and duration 20 ns, in the presence of a magnetic field of \(40~\mathrm{mT}\). $\mathrm{t}_0$ indicates the nucleation time of a seed domain, $\Delta \mathrm{t}$ is the transition time determined by the motion of the domain wall across the device, and $t'$ is the rest time until the end of the electric pulse. The switching trace is fitted to a Heaviside function that yields $t_0$ and $\Delta t$. (d) Transition time $\Delta t$, averaged over 250 independent acquisitions, at different external magnetic fields value, as a function of the current density.}
       \label{GdFeCoTRmeas} 
\end{figure*}

To test these predictions, we measured the SOT-induced domain wall dynamics in ferrimagnetic \pd{15 nm-thick Gd$_{29}$Fe$_{64}$Co$_{7}$},  Gd$_{30}$Fe$_{63}$Co$_{7}$, and Gd$_{31}$Fe$_{62}$Co$_{7}$ layers grown by sputtering onto a Pt (5 nm) underlayer (\textbf{Figure~\ref{GdFeCoTRmeas}a}). The composition of GdFeCo, a typical amorphous RE-TM ferrimagnet, was optimized to almost compensate the net magnetization and angular momentum close to room temperature \pd{(Supporting Information). We estimate that the Gd$_{30}$Fe$_{63}$Co$_{7}$ and Gd$_{31}$Fe$_{62}$Co$_{7}$ samples reach the compensation of the angular momentum above room temperature whereas the Gd$_{29}$Fe$_{64}$Co$_{7}$ sample is compensated below ambient conditions. Working in the vicinity of the minimum of the angular momentum} allows for achieving domain wall speeds in the range of km/s \cite{caretta2018fast, cai2020ultrafast, kim2017fast, sala2022asynchronous} \pd{and is necessary to observe relativistic dynamics (Supporting Information)}. Moreover, GdFeCo features a weak perpendicular anisotropy, large anomalous Hall resistance, and low Gilbert damping parameter \cite{hansen1991magnetic, kim2019low, bainsla2022ultrathin}. These characteristics make this alloy the ideal candidate to both \pd{trigger} relativistic effects and observe them through the time-resolved Hall effect measurements that are described next. 

The GdFeCo films were patterned into \textmu m-large large dots onto Pt Hall crosses by a combination of lithography and physical etching techniques (see \textbf{Figure~\ref{GdFeCoTRmeas}b}) \cite{salaSchematic, sala2022asynchronous}. The Pt Hall cross serves both as a current pulse injection line to generate the SOT that displaces the domain wall and detection probe of the real-time domain wall dynamics at sub-ns timescales. This temporal resolution is achieved by means of time-resolved measurements of the anomalous Hall effect. In this technique, ns-long electric current pulses induce SOT that switch the perpendicular magnetization of the GdFeCo dots and generate a transverse anomalous Hall effect. Because the latter changes proportionally to the out-of-plane magnetization, tracking the Hall voltage with an oscilloscope provides access to the magnetization dynamics, including the displacement of domain walls. More details on the device fabrication, experimental setup, and data analysis can be found in Ref. \cite{salaSchematic}.

Typical dynamics measured in this way comprise three phases whose duration we denote by $t_0$, $\Delta t$, and $t'$ (\textbf{Figure~\ref{GdFeCoTRmeas}c}). $t_0$ can be identified with the time required for nucleation of a seed domain at the device edge. Once this reversed domain is formed, the associated domain wall is displaced by SOT across the device. This process leads to a monotonic change in the Hall signal and defines the transition time \(\Delta \mathrm{t}\). After the domain wall has traveled across the entire device, the magnetization is completely reversed and remains steady for the time \(\mathrm{t}'\) until the electric pulse is over. The domain wall velocity can be derived by dividing the diameter of the dot by \(\Delta \mathrm{t}\), as confirmed by the linear scaling of the transition time with the diameter \cite{sala2023deterministic}. These dynamics can be measured with different combinations of the electric current inducing the SOT and in-plane magnetic field, which allows us to identify the regime of domain wall motion.

\begin{figure*}
     \centering
         \includegraphics[width = 0.9\textwidth]{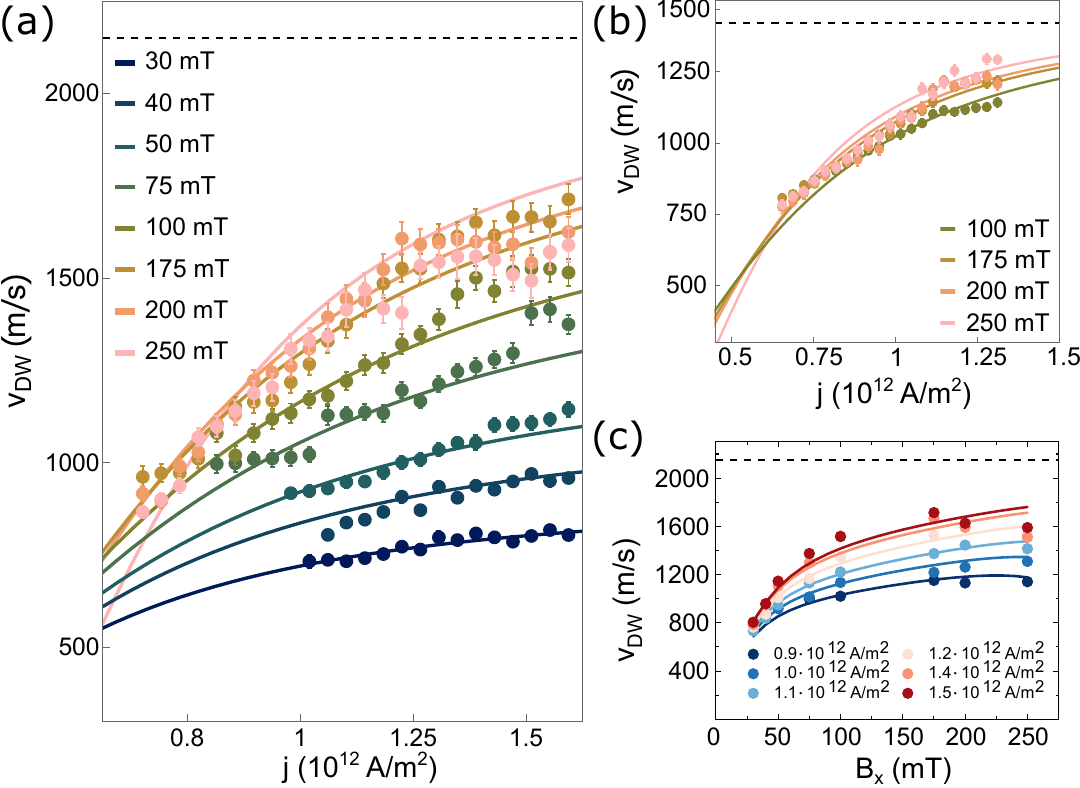}
    \caption{Domain wall speed as a function of the applied current density through the heavy metal layer for different in-plane magnetic fields in (a) 4 \textmu m-wide \(\mathrm{Gd_{30}Fe_{63}Co_7}\) and (b) 3 \textmu m-wide \(\mathrm{Gd_{31}Fe_{62}Co_{7}}\) dots. The lines are fits to Equation~\ref{rel2}. The horizontal dashed lines define the spin-wave group velocity estimated from the fits. (c) Domain wall speed as a function of the in-plane magnetic field at several current densities in the same device as in (a). The error bars in (a) and (b) are evaluated as the standard error of the mean of 250 measurements.}
       \label{Gd30} 
\end{figure*}

\textbf{Figure~\ref{GdFeCoTRmeas}d} shows the transition time \(\Delta \mathrm{t}\) measured in a 4 \textmu m-wide Gd$_{30}$Fe$_{63}$Co$_7$ dot. The transition time systematically decreases with increasing current density. This decrease results from the stronger SOT acting on the domain wall magnetization and is influenced by the magnetic field: larger fields result in a lower \(\Delta \mathrm{t}\). This behavior is indicative of the low-field regime in which the SOT and the restoring force of the magnetic field compete. Yet, further increasing the magnetic field brings the domain wall dynamics into a different regime. We observe that the data points start collapsing onto the same curve beyond a magnetic field of about 100 mT. This trend suggests the onset of a global field-independent saturation of the transition time. This interpretation is confirmed by the dependence of the domain wall speed on the current density shown in \textbf{Figure~\ref{Gd30}a} and \textbf{Figure~\ref{Gd30}b} for two different samples. Domain walls reach speeds higher than 1 km/s, in line with previous measurements of ultrafast dynamics in RE-TM ferrimagnets \cite{caretta2018fast, cai2020ultrafast}. However, domain wall velocities measured at magnetic fields higher than 100 mT cannot be distinguished from each other within the experimental uncertainties. Such an independence of the domain wall speed from magnetic field at $B_{\mathrm{x}} > 100$ mT is highlighted in \textbf{Figure~\ref{Gd30}c} and is reminiscent of the domain wall dynamics observed in ferrimagnetic garnets \cite{caretta1}. The high-field domain wall dynamics hint therefore at possible relativistic effects, \pd{which are confirmed in devices with different diameter (Supporting Information).}

To verify this possibility, we fitted the measured domain wall speed $v_\mathrm{DW}$ to Equation~\ref{rel2}. To reduce the number of free parameters, we measured the magnetic anisotropy energy density, the saturation magnetization, and the SOT efficiency (entering in $\Delta$ and $\tilde{H}_{\mathrm{DL}}$, Supporting Information) by combining magnetotransport and magnetometry techniques (see \textbf{Table \ref{paramextracted}}). The estimated values for these parameters fall in the typical range of RE-TM ferrimagnets \cite{hansen1991magnetic}. We also set the lattice spacing $d=4$~\AA \cite{hansen1991magnetic}. We left the remaining parameters free and estimated them via fits under the assumption that they do not change with the applied current, in line with Ref. \cite{caretta1} \pd{(Supporting Information)}. These parameters are: the antiferromagnetic exchange interaction $A$, the saturation magnetization of one sublattice $M_\mathrm{A}$, the Gilbert damping $\alpha$, the DMI $H_\mathrm{D}$, and the effective pinning field $H_\mathrm{pin}$. 

As shown in Figure~\ref{Gd30}a,b, the relativistic model reproduces well the experimental trends and captures the domain wall speed saturation at high magnetic fields. Note that the relativistic limit velocity lies above the experimental data points because, first, magnetic fields and current densities higher than those experimentally available would be necessary to saturate the domain wall speed to the spin-wave group velocity and, second, the limit is reached only asymptotically. The estimated spin-wave group velocity is \(\approx 2.1~\mathrm{km/s}\) for \(\mathrm{Gd_{30}Fe_{63}Co_7}\) and \(\approx 1.4~\mathrm{km/s}\) for \(\mathrm{Gd_{31}Fe_{62}Co_{7}}\). This difference \pd{may} arise from the distinct stoichiometry of the two samples, which, although being only 1\% different, can significantly affect the antiferromagnetic exchange interaction \cite{raasch1994exchange}. \pd{We do not observe clear relativistic effects in a GdFeCo sample with 29\% Gd concentration despite its domain wall speeds exceeding 1 km/s (Supporting Information). We attribute this difference to the difficulty of reaching the relativistic regime when working far from the compensation point of the angular momentum.} This sensitivity to the composition of the alloy \pd{and the use of moderate magnetic fields} may explain why relativistic dynamics have not been identified in previous studies of ultrafast domain wall dynamics in RE-TM ferrimagnets.

The other parameters extracted from the fits are similar for the two alloys \pd{with 30\% and 31\% Gd concentration} and generally match typical magnetic properties of RE-TM ferrimagnets (Table~\ref{paramextracted}). \pd{Their low and almost identical damping may contribute to the high velocities that we measure but does not explain the saturation of the speed}. We find a small pinning field on the order of a 30 mT and a negligible DMI. The fact that $H_\mathrm{D} \approx 0$ is supported by the observation that the velocity datasets \pd{and our fits} in Figure 3c converge to zero with \pd{negligible $B_{\mathrm{x}}$ intercept}.
\pd{The negligible role of the DMI is confirmed by magneto-optical Kerr measurements showing that domain walls mantain a straight profile when driven by current pulses (Supporting Information) and is in line with previous time-resolved imaging measurements performed in very similar devices \cite{sala2022asynchronous}. The weak DMI may result from the intermediate thickness of the GdFeCo layer, which hinders interfacial DMI effects without simultaneously allowing for bulk contributions \cite{kim2019bulk}. Effects associated with the domain wall elongation can also be excluded (Supporting Information). We note that our model does not directly take into account temperature variations caused by Joule heating, which makes all magnetic parameters time dependent on ns timescales. Despite this limitation, varying saturation magnetization and effective anisotropy by up to 40\% does not prevent us from capturing the speed saturation nor does it affect our estimate of the spin-wave group velocity (Supporting Information). This implies that, although our estimates may deviate from the actual time-dependent magnetic properties, our model and analysis are robust with respect to uncertainties on the parameters. That Joule heating, while unavoidable, cannot account for the speed saturation is additionally confirmed by Fig. 3c, where velocities measured with the same current density and hence the same thermal load reach a plateau at high magnetic field. Overall, these considerations and the} nice agreement between the measured dependence and the fits corroborate our interpretation of the domain wall saturation as the consequence of the relativistic limit set by the spin-wave group velocity. In contrast, the classical model cannot capture the experimental results because it does not account for the independence of the domain wall speed from the magnetic field in the high field regime. Attempting to fit the data to Equation~\ref{vsteadycomplete} yields a $\chi^2$ parameter that is 4 times smaller than that obtained with Equation~\ref{rel2}. We also note that the pinning of the domain wall by defects and inhomogeneities cannot explain the saturation of the domain wall speed because \pd{the pinning potential generally reduces as the strength of $H_{\mathrm{x}}$ increases (Supporting Information)}.

\section{Conclusion}
We have observed a saturation of the current-induced domain wall velocity in amorphous GdFeCo ferrimagnets under the action of in-plane magnetic fields. Our analysis proves that this saturation originates from the onset of relativistic dynamics in which the domain wall speed is limited by the spin-wave group velocity. The latter varies with the antiferromagnetic exchange interaction and thus depends on the RE-TM concentration. While domain wall dynamics have traditionally been studied in the low-field regime \cite{haltz2020measurement, caretta1}, where the maximum speed is governed only by the angular momentum compensation, our relativistic model shows that the ultimate speed is instead set by \pd{a more fundamental limit} -- the spin-wave group velocity -- independently of the applied current, magnetic field, or compensation of angular momentum. These results indicate that relativistic dynamics are not an exclusive feature of crystalline materials but are possible in amorphous magnets that are easy to integrate in spintronic devices. This work establishes a viable pathway to circumvent the integration challenges of crystalline ferrimagnets by leveraging amorphous ones as a scalable platform for ultrafast magnetic devices operating at the ultimate speed limit.




\medskip
\textbf{Acknowledgements} \par 
We thank Pietro Gambardella for supporting the project. P.D., L.M. and S.A. acknowledge funding from IIT Genova. 
G.S. acknowledges the support from the Swiss National Science Foundation (grant PZ00P2\_223542).

\medskip
\textbf{Conflict of Interest} \par 
The authors declare no conflict of interest.

\medskip
\textbf{Data Availability Statement} \par 
The data and the code that support the ﬁndings of this study are available on Zenodo with doi 10.5281/zenodo.16795567.

\medskip

%
\bibliographystyle{MSP}
\bibliography{bibliography}




\clearpage
\newpage

\begin{table}[h!]
 \caption{Parameters of Equation~\ref{rel2} for two alloy compositions. A, $M_{\mathrm{A}}$, $\alpha$, $H_{\mathrm{pin}}$ are fitted to the experimental data. $v_{\mathrm{g,max}}$ is derived from the fitted parameters. $K_{\mathrm{eff}}$, $M_{\mathrm{S}}$ and $\theta_{\mathrm{DL}}$ are experimentally measured.}
    \begin{tabular}[htbp]{@{}lll@{}} 
    \hline
    Parameters & $\mathrm{Gd_{30}Fe_{63}Co_{7}}$ & $\mathrm{Gd_{31}Fe_{62}Co_{7}}$\\
    \hline\hline
    $A$  & $\mathrm{1.9 \,pJ/m}$ & $\mathrm{1.2 \, pJ/m}$\\
    \hline
    $M_\mathrm{A}$ & $\mathrm{4.3\cdot 10^{5} \,A/m}$ & $\mathrm{4.1\cdot 10^{5} \, A/m}$\\
    \hline
    $\alpha$ & 0.006 & 0.005\\
    \hline
    $H_\mathrm{pin}$ & 28  mT & 29 mT\\
    \hline
    $v_\mathrm{g,max}$ & $\mathrm{2.1 \, km/s}$ & $\mathrm{1.4 \, km/s}$\\
    \hline\hline
    $K_\mathrm{eff}$ & $\mathrm{9.2\cdot 10^3 \frac{J}{m^3}}$ & $\mathrm{10\cdot 10^3 \frac{J}{m^3}}$\\
    \hline
    $M_\mathrm{S}$ & $\mathrm{4.3\cdot 10^4 \frac{A}{m}}$ & $\mathrm{3.8\cdot 10^4 \frac{A}{m}}$\\
    \hline
    $\theta_\mathrm{DL}$ & $0.13$ & $0.13$\\
    \hline
    \end{tabular}
    \label{paramextracted}
\end{table}

\end{document}